\documentclass[structabstract]{aa}

\usepackage {txfonts}
\usepackage{graphicx}
\usepackage{natbib}
\bibpunct{(}{)}{;}{a}{}{,} 

\begin{document}

\title {GRB 050502B optical afterglow: a jet break at high redshift\thanks {Based on observations
made with the Italian Telescopio Nazionale Galileo (TNG) operated on the island of
La Palma by the Fundaci\'on Galileo Galilei of the INAF (Istituto Nazionale di
Astrofisica) at the Spanish Observatorio del Roque de los Muchachos of the Instituto
de Astrof\'{\i}sica de Canarias (program AOT11-59) and with ESO Telescopes at the La
Silla Paranal Observatories under program ID 177.A-0591.}}

\author{P. Afonso\inst{1}
\and J. Greiner\inst{1}
\and E. Pian\inst{2,3,4}
\and S. Covino\inst{5}
\and D. Malesani\inst{6}
\and A. K\"upc\"u Yolda\c{s}\inst{4}
\and T. Kr\"uhler\inst{1,7}
\and C. Clemens\inst{1}
\and S.\,~McBreen\inst{1,8}
\and A. Rau\inst{1}
\and D. Giannios\inst{9}
\and J. Hjorth \inst{6}}

\offprints{P. Afonso, \email{pafonso@mpe.mpg.de}}

\institute{Max-Planck-Institut f\"ur extraterrestriche Physik, Giessenbachstra\ss e 1, 85748 Garching, Germany
\and Osservatorio Astronomico di Trieste, Istituto Nazionale di Astrofisica, Via G.B. Tiepolo 11, 34131 Trieste, Italy
\and Scuola Normale Superiore, Piazza dei Cavalieri 7, 56126 Pisa, Italy
\and European Southern Observatory, Karl-Schwarzschild-Stra\ss e 2, 85748 Garching, Germany
\and Osservatorio Astronomico di Brera, Istituto Nazionale di Astrofisica, Via Brera 28, 20121 Milano, Italy
\and Dark Cosmology Centre, Niels Bohr Institute, University of Copenhagen, Juliane Maries Vej 30, 2100 Copenhagen \O, Denmark
\and Universe Cluster, Technische Universit\"at M\"unchen, Boltzmannstra\ss e 2, 85748 Garching, Germany
\and University College of Dublin, Science Centre, School of Physics, Belfield, Dublin 4, Ireland
\and Princeton University, Department of Astrophysical Sciences, Peyton Hall, Princeton, NJ 08544-1001, USA
}

\date{}

\abstract {} {GRB\,050502B is well known for the very bright flare displayed in its X-ray light curve. Despite extensive studies, \mbox {however}, the optical light curve has never been discussed and its redshift is unconstrained. Possible correlations between optical and X-ray data are analysed.} {Photometric data from TNG in the $R$ and $I$ bands were used to compare the optical afterglow with the X-ray light curve. The $HyperZ$ package and a late time VLT host observation were used to derive redshift estimates.} {The $I$-band afterglow decay followed a power-law of index $\alpha = 2.1\pm 0.6$, after a late break at $\sim 1.3\times 10^{5}$ s. The $R-I$ color is remarkably red and the broadband spectral index $\beta_{\rm OX} = 0.9 \pm 0.1$ is consistent with the X-ray spectral slope $\beta_{\rm X}$. Although a photometric redshift of $z>4$ is the most conservative result to consider, a photometric redshift of $z = 5.2\pm 0.3$ is suggested with no extinction in the host, based on which an isotropic energy $E_{\gamma,\rm iso} =(3.8\pm {0.7}) \times10^{52}$\ erg and a jet opening angle $\theta\sim 3.7\degr$ are subsequently derived.} {The combined X-ray and optical data suggest an achromatic break, which we interpret as a jet break. The post jet break slope obeys roughly the closure relation for the jet slow cooling model. Because of the afterglow's very red color, in order for the redshift to be low ($z<1$), extinction, if present in the host, must be significantly high. Since the optical-to-X-ray index is consistent with the X-ray spectrum, and there is no XRT evidence for excess $N_{\rm H}$, GRB\ 050502B was likely at high redshift.}



\maketitle

\section{Introduction}
The gamma-ray burst (GRB) 050502B was a remarkable event showing a conspicuous X-ray flare lasting from $\sim$ 400 to 1400 s after trigger, with as much energy ($ \sim 9 \times 10^{-7}$ erg\ cm$^{-2}$, in the 0.3-10.0\ keV band) as the prompt emission itself [($8.0\pm 1.0) \times 10 ^{-7}$ erg cm$^{-2}$, in the 15-350\ keV band]. The X-ray light curve also showed later activity and a second less intense flare, followed by a possible jet break at $\sim1.1 \times 10^{5}$\ s \citep{Falcone,Burrows}. Given the exceptionality of this GRB, it is important to study its optical afterglow, searching for properties that may lead to a better understanding of its phenomenology. 



As expected, when the relativistic jet slows down, the aberration of light due to Special Relativity effects becomes less important, and thus the beaming angle, $\theta_{\rm b} \sim 1/\Gamma$, increases within the jet \citep {Rhoads}. A jet break occurs when the Lorentz factor is such that $1/\Gamma > \theta_{\rm jet}$, resulting in a significant decrease of the afterglow brightness and steepening of the light curve at all wavelengths. In the \textit{Swift} satellite era \citep{Gehrels}, X-ray light curves are much better sampled than previously, but simultaneous optical and X-ray light curve breaks have been rarely observed. For example, from the sample of \citet{Liang}, with \textit{Swift} X-Ray Telescope \citep [XRT;] [] {Burrows05} data for 179 GRBs and 57 pre- and post-\textit{Swift} GRBs optical afterglow data, only 7 of them show an achromatic break. This apparent lack of jet breaks is still in debate and could be due to a combination of higher redshifts, fainter bursts, larger opening angles, etc. Models that go beyond the fireball external shock afterglow scenario have also been explored, as they include an extra component that can be described as ``late prompt" emission due to late time activity of long living central engines. As discussed by \citet{Ghisellini} and \citet{Nardini}, the lack of achromatic breaks could be explained by the presence of the ``late prompt" emission.     

The analysis of GRB 050502B optical data from the 3.58\,m Telescopio Nazionale Galileo (TNG) was done with focus on correlations between the optical and X-ray afterglow behavior. In the next pages the X-ray light curve is discussed first, following \citet{Falcone} as a guideline. Evidence for a jet break and large redshift are presented later in $\S\ 3$.

\section{Observations and data reduction}

\subsection{Optical data from TNG and VLT}

At 09:25:40 UT, the Swift Burst Alert Telescope \citep [BAT;] [] {Barthelmy} triggered and located GRB050502B \citep {Pagani}. The XRT repository indicates that the afterglow of GRB 050502B was observed at the coordinates RA = 09:30:10.11, Dec = +16:59:47.9 (J2000.0), being located within the XRT UVOT-enhanced position 1.4 arcsecond radius error circle. The detailed TNG $I$ and $R$ bands data, taken on 2005, May 3, 4 and 5 (1, 2 and 3 days after the trigger, respectively), are summarized in Table 1. Raw images were corrected for bad pixels, debiased and flat-fielded in a standard way with IRAF \citep{Tody}. In addition, the TNG $I$-band images severely affected by fringing were corrected using standard routines. Photometry was done also with IRAF, using the USNO-B1.0 catalog for calibration purposes, which has a systematic error of $\pm 0.2$\,mag. In order to calibrate the afterglow photometry, 40 to 50 stars were used per final science image. The TNG photometric uncertainties listed in Table 1 are statistical only. Later in time the Sloan Digital Sky Survey (SDSS) Data Release 6, finally covered the field of view of GRB\,050502B. A cross-calibration between USNO-B1.0 and SDSS catalogs yielded consistent zero points. Conversion from $r$ to $R$-band and $i$ to $I$-band was performed using observations of Landolt standard stars \citep {Landolt} in SDSS fields, following \citet {Chonis}.


 

To compute the flux density for the $R$ and $I$ bands, Vega flux reference values, $f_{0}$, were taken from \citet{Fukugita}. The correction for Galactic extinction was based on the interstellar extinction curves derived by \citet{Cardelli} and \citet {O'Donnell}, with $E(B-V) = 0.03$ mag \citep{Schlegel}.

Afterglow spectroscopy was attempted at the European Southern Observatory (ESO) Very Large Telescope (VLT), but unluckily the weather conditions did not allow for meaningful observations. Finally, late-time observations of the field of GRB\,050502B were taken in the $R$-band with the Focal Reducer and Spectrograph 2 (FORS2) instrument at VLT, within the large program on GRB host galaxies (177.A-0591; PI: Hjorth). Observations were carried out 321 days after the GRB (on March 23, 2006), for a total exposure time of 2000~s. At the afterglow position, no object is detected down to a $3\sigma$ limiting magnitude of $R > 26$\ mag.

\subsection {Other optical data and upper limits}

All published optical data are presented in Table 1.
These include several upper limits (ULs) and in particular one more detection further to those by TNG, reported by the Australian National University (ANU) 1-m telescope \citep{Rich,Cenko}. Unfortunately the ANU data were lost on a corrupted disk, being impossible to produce further refined photometry to include in this paper (B. Schmidt, personal communication).

During the giant X-ray flare, the \textit{Swift}-Ultraviolet and Optical Telescope \citep [UVOT;] [] {Roming} did not detect any optical counterpart, reporting only UL for a $9 \times 10$ s exposure \citep{Schady}. The UVOT repository \citep {Roming2} gives a detailed sequence of observation times, as each of the $UBV$ filters was used, and more than the initial $9 \times 10$ s exposure were taken. The time error bars in Fig. 1 reflect these data. For the time period of the X-ray flare, observations with the $U$ filter started 192 s after the burst, ending 960 s after, with interruptions to switch to other filters. For the $V$ and $B$ filters the start and end times are 234-920 s and 290-975 s, respectively. Table 1 indicates the median times for these UVOT observations. Several subsequent UVOT observations, roughly following the timing of XRT exposures, continued late after the flare but resulted only in shallow upper limits (not reported in Table 1).

\begin{table}
\caption{Data from TNG and other telescopes taken from the literature. The first column gives the mid time or the start of observations. For the ART and Himalayan telescopes data, the UL calibration catalog was not mentioned; thus Vega magnitudes were assumed and converted to AB magnitudes here.}            
\label{table:1}      
\begin{center}       
\renewcommand{\footnoterule}{}   
\begin{tabular}{c c c c c}        
\hline\hline                 
Time (s) & Filter & Telescope & Exposure (s) & Magnitude (AB) \\     
\hline                 
56.7 & $B$ & $\rm Ashra\ 0.3m^{a}$ & 1.4 & $>15.9$\\
56.7 & $R$ & $\rm Ashra\ 0.3m^{a}$ & 1.4 & $>14.7$\\
180 & $B$ & $\rm Ashra\ 0.3m^{a}$ & $25\times4$ & $>18.2$\\
180 & $R$ & $\rm Ashra\ 0.3m^{a}$ & $25\times4$ & $>17.1$\\
576 & $U$ & $\rm UVOT^{b}$ & $9\times10$ & $>19.7$\\
577 & $V$ & $\rm UVOT^{b}$ & $9\times10$ & $>18.4$\\
633 & $B$ & $\rm UVOT^{b}$ & $9\times10$ & $>19.1$\\
570 & $V$ & $\rm ANU\ 1m^{c,d,e}$ & not given & $>20.6$\\
900 & $V$ & $\rm ANU\ 1m^{c,d,e}$ & not given & $>20.1$\\
1080 & $V$ & $\rm ANU\ 1m^{c,d,e}$ & not given & $>20.7$\\
1200 & $V$ & $\rm ANU\ 1m^{c,d,e}$ & not given & $>21.7$\\
3950 & $I$ & $\rm ANU\ 1m^{c,d,e}$ & $\rm total \sim4020$ & $20.23\pm 0.09$ \\ 
7339 & $I_{\rm c}$ & $\rm ART\ 0.36 m^{f}$ & $87\times60$ & $>18.9$ \\
15560 & $R$ & $\rm Lulin\ 1m^{g}$  & not given & $>21.9$ \\
19160 & $R$ & $\rm Himalayan\ 2m^{h}$  & $4\times600$ & $>22.6$ \\
22566& $V$ & $\rm Aries\ 1m^{i}$  & $12\times300$ & $>21.4$ \\
128796 & $I$ & TNG & 60 & $21.73\pm 0.21$ \\
129104 & $I$ & TNG & 300 & $21.28\pm 0.08$ \\
129521 & $I$ & TNG & 300 & $21.15\pm 0.07$ \\
157680 & $i$ & $\rm P60^{c}$ & $10\times120$ & $>21.3$\\
215661 & $R$ & TNG & $11\times300$ & $23.71\pm 0.20$ \\
219235 & $I$ & TNG & $2\times300$ & $22.59\pm 0.12$\\
302676 & $I$ & TNG & $8\times300$ & $>23.7$\\ 
$\sim 2.8\times 10^{7}$ & $R$ & VLT & 2000 & $>26$\\
\hline                                    
\end{tabular}
\end{center}
$^a$\citet{Sasaki},$^b$\citet{Schady}, $^c$\citet{Cenko},$^d$\citet{Rich},
$^e$\citet{Rich05}, $^f$\citet{Torii}, $^g$\citet{Sanchawala},$^h$\citet{Bhatt},
$^i$\citet{Misra}.
\end{table}

\begin{figure*}
\centering
  \includegraphics[width=17cm]{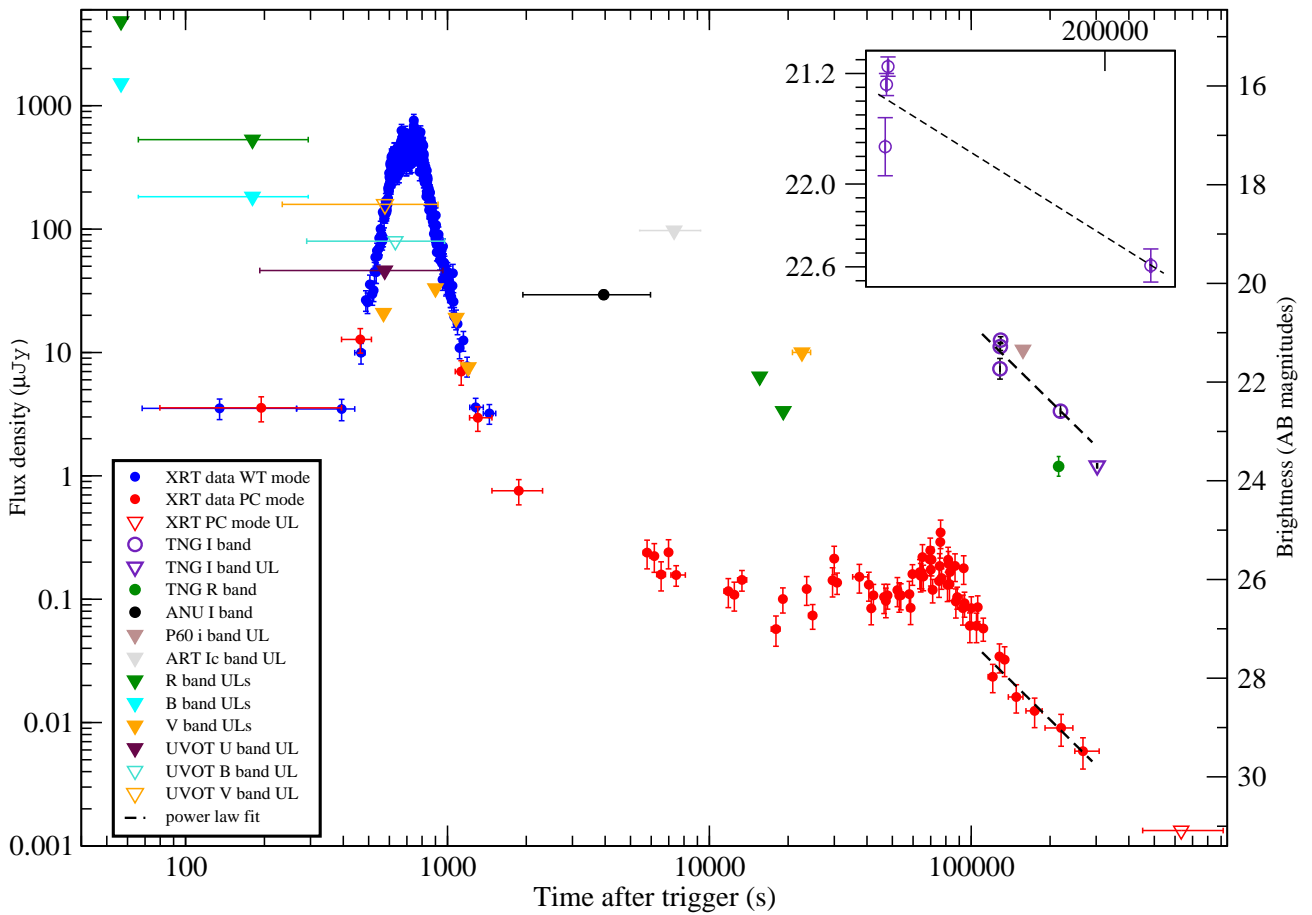}
    \caption{Optical and X-ray (1 keV) light curves. Horizontal error bars represent the exposure times.  Dashed lines represent the time power-law fits to the 4 TNG $I$-band detections and to the last 7 X-ray data points. The inserted box zooms in on the TNG $I$-band detections.}
     \label{Fig.1}
\end{figure*}


\section{Results}
\subsection{Optical and X-ray light curves: the jet break}

To produce the X-ray light curve shown in Fig. 1, in order to compute the X-ray flux densities at 1\ keV, the unabsorbed flux integral was computed over the energy range 0.3-10\ keV. The required spectral indices $\beta$ are fitted for the window timing (WT) and photon counting (PC) XRT modes and taken from the \textit{Swift}-XRT spectrum repository \citep{Evans}.
 
The main feature of the XRT light curve is a giant flare that occurs on top of an underlying decaying afterglow, obeying a power-law $ F(t)\propto t^{-\alpha}$ of time index $\alpha = 0.8\pm 0.2$. Though not shown anymore in the current light curve found in the XRT repository, \citet{Falcone} report for their binned data that the pre-flare decay rate is maintained in the post-flare, up until to $\sim 10^{4}$\,s after the GRB start time. The light curve then exhibits a flattening combined with two over-imposed rebrightenings at $ \approx 3 \times 10^{4}$ and $ \approx 8 \times 10^{4}$ s, respectively. The latter marks the onset of a steep decline, which could be a jet break. From the shallow decay phase the light curve evolves directly into post-jet break, not showing the less steep pre-jet break phase \citep {Panaitescu}.


	

Considering the available $I$-band data at a time \mbox{$t > 1.28\times 10^{5}$\,s}, when fitted with a power-law, the time decay index is $\alpha = 2.1\pm 0.6$. This late time decay must be a post-break decay, since the ANU telescope detection appears to constrain the early time decay to a slope shallower than that of the TNG.

When fitting the last seven X-ray time bins with a power-law, the result is $\alpha = 2.1\pm0.3$. The first of these seven bins matches the start of our optical observations at $\sim 1.28 \times 10^{5}$\ s. If considering one extra earlier bin, i.e., for\, $t> 1.1 \times 10^{5}$\ s (like in \citealp {Falcone}), the time index is then $\alpha = 2.3\pm0.3$, while \citet{Falcone} got a power-law time index of $\alpha = 2.8\pm 0.8$.

The results presented in this work are compatible with those from \citet {Falcone}, despite using a slightly different X-ray data binning. From the currently available \textit{Swift}-XRT data repository, it can be established that for the last seven bins the count rate light curve has a minimum of 17 and a maximum of 99 counts per bin. Thus such binning is statistically meaningful. 


Noting that the steep optical decay overlaps in time and has a time index consistent (within errors) with that in the X-rays, the data suggest thus a jet break.

The spectral index in any given band across a jet break should be unchanged. Therefore, in order to test the jet break hypothesis, an averaged spectrum was created in the \textit{Swift}-XRT repository for all points with $t>1.1 \times 10^{5}$\ s, resulting in a spectral index of $\beta_{\rm X}= 1.0_{-0.2}^{+0.4}$, with a best fit value in excess of the Galactic hydrogen column density of $N_{\rm H}= 0_{-0}^{+3.7} \times 10^{20}$ cm$^{-2}$. For the interval [$\sim 1000$, 20000] s, the spectral index is basically the same, with $\beta_{\rm X}= 1.0_{-0.2}^{+0.3}$. 

\subsection{The Spectral Energy Distribution}

The broadband spectral energy distribution (SED) was computed at the epoch of our single $R$-band detection. Since the $R$-band point precedes the closest $I$-band point by $\sim$ 1 hr, the $I$-band post-jet break slope was used to extrapolate the $I$-band value (0.04 magnitudes brighter) at the $R$-band observation time. The effective wavelengths, $\lambda_{0}$, corresponding to the $R$ and $I$ bands, were taken from \citet{Fukugita}. The X-ray data point used in the SED has a flux density (at 1 keV) of $\ 9\times 10^{-3}\ \mu$Jy, at the time of the penultimate bin point in the X-ray light curve. This X-ray bin point matches the extrapolated $I$-band and observed $R$-band observation time.

Fitting the broadband SED with a simple $F_{\nu}(\nu)\propto \nu^{-\beta}$ power-law yields a spectral index  $\beta_{\rm OX} = 0.9 \pm 0.1$.
This value is consistent with the X-ray spectral index ($\S\ 3.1$), thus there is no indication of a break between the optical and X-ray spectra and the SED between the $I$-band and $\sim 5$ keV is well described by a single power-law with $\beta_{\rm OX} = 0.9 \pm 0.1$.


Finally, although there is only one epoch at which the broadband SED could be computed, the temporal decay indices in the optical and X-rays are consistent as well, which is a necessary condition to keep having $\beta_{\rm X}$ consistent with $\beta_{\rm OX}$ in other epochs. In fact if the optical and X-ray decay indices become different, it would imply necessarily a spectral break between optical and X-rays.

\subsection{The closure relations}

Considering the fireball synchrotron shock model as the major radiation mechanism responsible for the afterglows (e.g., \citealp{Zhang}) and following, e.g., \citet{Sari}, GRB 050502B roughly obeys the closure relations for a model of a uniform jet (valid for both ISM and wind cases) in a typical slow cooling regime, for $\nu>\nu_{\rm c}$, where $\nu_{\rm c}$ is the synchrotron cooling frequency. The electron power-law index, $p$, is deduced from the temporal decay index at the post-jet break phase, i.e., $F_{\nu} \propto\ t^{-p}$. Thus for $p\sim 2.1$ and $\beta_{\rm OX} = 0.9\pm 0.1$, within errors, it is observed $\alpha = 2\beta$.

\subsection{Photometric redshift}

The quasi-simultaneous $R$ and $I$ bands TNG observations at 215.6 and 219.2 ks, respectively, result in $R-I=1.12\pm 0.23$\ mag. The correction for simultaneity (using the time index computed for the $I$- band) is smaller than the error due to the uncertainty in the photometry, resulting in $R-I= (1.16 \pm 0.23)$\ mag. This very red color hints for either high redshift or high extinction in the host, or both. These possibilities were explored using the $HyperZ$ photometric redshift code from \citet{Bolzonella}. No reddening in the host and a fixed $\beta_{\rm OX} = 0.9\pm 0.1$ produce the best fitting result of $z = 5.2 \pm 0.3$ (see Fig. 2). Since $\beta_{\rm OX}$ and $\beta_{\rm X}$ are consistent, this further favors the unextinguished model.




\subsection{Burst energetics}

According to \citet{Cummings} the fluence in the BAT range (15-350\ keV) is $F_{\gamma} = (8.0\pm 1.0) \times 10^{-7}$ erg cm$^{-2}$. To estimate the energy of the burst if occurring at $z= 5.2$, the corresponding luminosity distance is computed in a concordance cosmology model \citep{Spergel}. With $H_{0} = 71$ km\ s$^{-1}$ Mpc$^{-1}$, $\Omega_{\rm M} = 0.27$ and $\Omega_{\Lambda} = 0.73$, it results $D_{\rm L} = 49.84$\ Gpc.  For the BAT range of energies, considering the errors in the \mbox{fluence} and redshift, this results in an isotropic-equivalent energy of $E_{\gamma,\rm iso}= 4\pi D_{\rm L}^{2} F_{\gamma} (1+z)^{-1} = (3.8\pm 0.7)\times 10^{52}$ erg, which is a typical value for long GRBs.



The jet opening angle $\theta$, which is dependent on the redshift and jet break time, can also be computed. To compare our results with $\theta \sim 8.1\degr$ from \citet{Falcone}, the same formula is followed, other than a small correction, required for dimensionality consistency, where $n_{0} = n_{\rm ism}/ (1\ \rm cm^{-3})$. Thus where \citet{Falcone} assumed $z = 1$ and $E_{52}=1$, in  $\theta = 7.8\degr t_{5}^{3/8}\ E_{52}^{-1/8}\ n_{0}^{1/8}\ [(1+z)/2]^{-3/8}$, it is now $t_{5} = 1.1$, for a jet break time of $\sim 1.1 \times 10^{5}$ s, $E_{52} = 3.8 \eta_{\gamma}$,
$n_{0}\sim 1$, for a uniform external interstellar medium density, and $z = 5.2$; this gives  $\theta\sim 3.7\degr$. Note that, as in other works (e.g., \citealp {Ghirlanda07, Frail}), $\eta_{\gamma} = 0.2$ corresponds to an assumed radiative efficiency of 20\%. Given that the bolometric $E_{\gamma,\rm iso}$ must be larger than that observed within the BAT range used here, $\theta\sim 3.7\degr$ is thus an upper limit, within the typically observed jet opening angles \citep{Ghirlanda07}.

\section{Discussion}

\subsection{Optical and X-ray light curves: the jet break}

The second and last bump on the X-ray light curve masks the true start of the jet break, while in the optical a possible bump or small flare cannot be ruled out. There is a small increase in brightness in the first three $I$-band data points, but it is not significant given that the first point does not have the best S/N. Also the second and third $I$-band data points are consistent, within errors, with no flaring, as can be better seen in the zoomed inserted box in Fig. 1. In any case even if this was a flare, in order to explain the rapid decay from the first 3 detections in May 3 (starting at 128 796 s after the trigger) to the following detection in May 4 (at 219 235 s), occurring simultaneously in the X-rays, the jet break still seems the best answer.

It is interesting to compare the last X-ray bump with the complex light curve of GRB 071010A \citep {Covino}, where after a rebrightening at 0.6 days after the trigger, a steeper decay began both in the X-ray and optical, as evidence of a jet break. In this and other GRBs with similar light curves, there is no plateau phase prior to the bump though - and rebrightening explanations related to the density profile of the circumburst medium have been invoked (among others).



The optical post-jet break slope could have been further constrained if earlier data had been taken, since the fitting of the X-ray jet break data points starts some $\sim 33$  min\ (2000\ s) earlier than our first optical data point. Note also that the TNG $I$-band non-detection at the significance of $2 \sigma$ (i.e., the UL of May 5, 2005, at $\sim 302$ ks post-burst) favors an even steeper post-jet break.

\subsection{Photometric redshift}

When considering extinction in the host, with data from only two filters, it is impossible to strongly constrain both the redshift and $A_{\rm V}$, given that they are generically anticorrelated, as shown in Fig. 2. Several host extinction models were used with $HyperZ$ and in Fig. 2 the Small Magellanic Cloud (SMC) model is shown as an example to illustrate the dependence with redshift. This suggests that a low redshift is allowed for large $A_{\rm V}$ values.

As noted by \citet{Falcone}, during the observations the total $N_{\rm H}$ can be considered constant at approximately the \citet{Dickey} value for the Galactic hydrogen column density, $N_{\rm H} = 3.6 \times 10^{20}$ cm$^{-2}$. Constructed with data from the XRT repository, a time-averaged spectrum between $T_{0} + 68$ and $\sim 1.4\times 10^{5}$ s is best fitted (assuming $z = 0$) with an absorbed power-law model where \mbox{ $N_{\rm H}= 8.2_{-8.2}^{+20.6} \times 10^{19}$ cm$^{-2}$}. This is a typical intrinsic excess value, similar to those observed in the high-redshift GRBs sample of \citet {Grupe}. Within the XRT fit error this result is consistent with zero, but taking the upper error value allows to compute an upper limit on the possible excess extinction.

There is no exactly known relation between $A_{\rm V}$ and $N_{\rm H}$ in GRB environments. To find the dependence of the $A_{\rm V}$ ULs with $z$, the X-ray spectral fitting package \textit{Xspec} \citep {Arnaud} was used, placing the excess $N_{\rm H}$ column at redshifts of 0, 1, 2, 3, 4 and 5. The cubic polynomial that represents the XRT UL curve in Fig. 2 fits the data points obtained from the Milky Way based \citet{Predehl} relation, where $N_{\rm H} = 1.79\ A_{\rm V}\times 10^{21}$\,cm$^{-2}$. Although, in general, high $N_{\rm H}$ corresponds to high $A_{\rm V}$, the dust-to-gas ratio in GRB hosts can be different and normally is smaller than in the Milky Way. \citet {Galama} report some examples with high $N_{\rm H}$ but with $A_{\rm V}$ being 10-100 times smaller than expected. Likewise the sample of \citet {Stratta} favors a dust-to-gas ratio of $\sim 1/10$ that of the Milky Way, as in a SMC environment. Therefore a more reasonable approach was to scale down by a factor of 10 the XRT UL data points presented in Fig. 2, after being computed from the \citet {Predehl} relation.


Stressing that all the \textit{Xspec} results are compatible with zero extinction, an inspection of Fig. 2, e.g., for $z=4$ at the $1 \sigma$ level, shows nevertheless an UL of $A_{\rm V} \sim 0.5$\,mag. This is a relatively high extinction, that cannot be probed directly at the rest frame by XRT, due to its spectral coverage. So a combination of both high redshift and $A_{\rm V}$ can also not be excluded and a more conservative result of $z>4$ should be considered.

The GRB $A_{\rm V}$ distribution is poorly constrained. While in the past samples were biased towards low extinction bursts, showing typical intrinsic reddening to be well below $A_{\rm V}\sim 0.5$ mag \citep {Kann}, recent observations provide evidence for highly reddened afterglows \citep {Watson06, Rol, Kruehler, Eliasdottir, Prochaska}. The case of GRB 070306, detected initially in the $K$-band, is a remarkable example of a highly reddened burst with $z = 1.496$ and suggested $A_{\rm V} = 5.6$ mag \citep {Jaunsen}. \citet {Tanvir} report GRB 060923A as being \mbox{another similar} extreme case.

As an additional note, since dust affects mostly the rest frame UV, normally a somewhat higher redshift ($\sim 2-3$) is needed in order to get such $R-I$ red color. Furthermore in the sample of \citet {Schady07} all highly reddened bursts due to intrinsic extinction have large $N_{\rm H}$ in the XRT data, which is the opposite of the present case.

For $z > 4.5 $, UVOT cannot detect anymore the Lyman limit \citep {Roming} and the UVOT data do not indicate the presence of a strong rebrightening in the optical simultaneous to the giant X-ray flare. Note however that, independent of its origin, the very red color only strictly suggests very little flux in the blue and UV bands probed by UVOT, hence the occurrence of a dimmer optical flare cannot be totally excluded.


If allowing for the possibility that there is no correlation between $A_{\rm V}$ and $N_{\rm H}$ (e.g., \citealp{Jakobsson}, \citealp {Watson}), the constraint from the X-ray spectrum could be circumvented. In this case, however, the extinction corrected optical fluxes would be increasingly above the 0.9 power-law extrapolation from the X-ray band, thus producing a convex SED. Such an SED shape has not been proposed within any afterglow model, was never observed and seems therefore an extremely unlikely option.

Regarding the $R$-band VLT data, since GRB host galaxies can be very faint (e.g, \citealp{LeFloc'h}) the VLT non-detection of a host galaxy provides only a mild limit on the redshift. It is difficult to quantify a lower limit given the scarcity of data, but looking into samples of GRB host galaxies, as in those of \citet{Wainwright} and \citet{Savaglio}, hosts with magnitude $R \gtrsim 26$ typically have $z \gtrsim 1.5$. This can be taken as a rough lower limit to the redshift of GRB\,050502B \citep {Malesani}.




\begin{figure}
\centering
  \includegraphics[width=8.2cm]{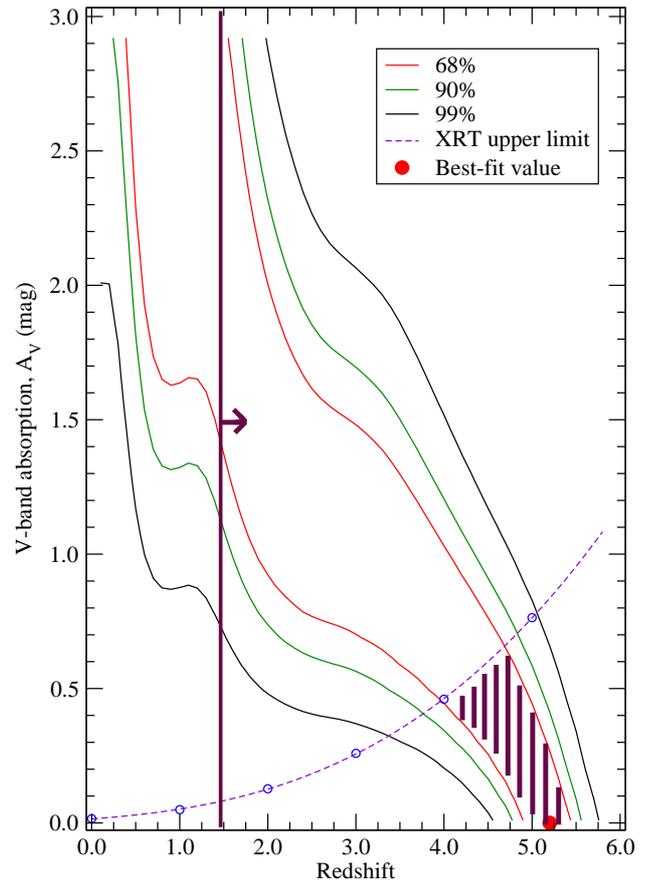}
    \caption{Intrinsic reddening vs. redshift dependence, for an SMC extinction curve with SED slope fixed to $\beta = 0.9$. The 68\%, 90\% and 99\% confidence contours are plotted. The XRT-fit line represents the $A_{\rm V}$ upper limits, converted as described in the text. The fit errors are, though, compatible with zero intrinsic $N_{\rm H}$. The maroon shaded area represents the $1\sigma$ region of the contour plot constrained by the XRT upper limits. The vertical line at $z = 1.5$ represents the VLT lower limit due to the lack of a host detection.}
     \label{Fig.2}
\end{figure}

\subsection{Burst energetics}


In $\S\ 3.5$, $E_{\gamma,\rm iso}$ was computed for $z = 5.2$. Just for reference, if taking $z=1$, then  $E_{\gamma,\rm iso}= 2.1 \times 10^{51}$ erg, which compared to samples of other bursts with known redshift (e.g., \citealp{Amati}), would put GRB 050502B to the lower energy end of GRBs. 

A hard question to answer is how to compute a bolometric rest frame isotropic-equivalent energy with a proper k-correction without knowing exactly the value of the peak energy, $E_{\rm p}$. Following \citet{Cummings} and \citet {Sakamoto08}, the BAT spectrum (range 15-350\ keV) is well fitted by a single power-law with a photon index $\Gamma = 1.59 \pm 0.14$. From the BAT spectrum, $\sim 90$\ keV is roughly the highest energy at which a reasonable detection still occurs and, given the value of\, $\Gamma$, this could be assumed as a lower limit for $E_{\rm p}$. An empiric estimator could be considered instead, as in \citet{Sakamoto}, where $ \log (E_{\rm p}/1~\rm keV) = 3.258 -0.829\ \Gamma$. Combining the estimator's own 1 $\sigma$ uncertainties and those in $\Gamma$, results then in $E_{\rm p} = 85_{-52}^{+184}$ \ keV. Taking $z =5.2$, 2.5 and $\Gamma$ as the high and low energy indices, respectively, of a Band function spectral shape \citep{Band}, if considering only the uncertainties in $\Gamma$, results already in a large $\sim 200$\ keV uncertainty range for the rest frame peak energy, $E_{\rm p,i} = 530_{-90}^{+110}$\ keV. These estimators are only valid for BAT spectra that can be fitted by a single power-law, but older empiric estimators produce significantly different results (e.g. \citealp{Liang07}).


In order to check if GRB\ 050502B obeys the Amati and Ghirlanda relations, the ``standard'' isotropic-equivalent bolometric energy release in the rest frame range of 1 keV to 10 MeV needs to be computed. It is difficult to reach firm conclusions given the errors in $F_{\gamma}$, $\Gamma$, $z$ and the uncertainties in $E_{\rm p}$ and in the Band function indices. In a time dilation and k-correction computation \citep{Amati02}, different combinations of parameters produce significantly different results. For reference, an exercise within the parameters errors, say with $F_{\gamma} = 9\times 10^{-7}$\,erg\,cm$^{-2}$, $z= 5.5$, Band energy indices $\beta = 2.5$, $\alpha = 1.7$ and $E_{\rm p}= 70$\,keV, produces a very reasonable $E_{\rm p,i}= 455$\,keV and $E_{\gamma, \rm iso} = 9.3 \times 10^{52}$\,erg. Following one of the latest GRB samples to test the Amati relation \citep{Amati}, this result places GRB\ 050502B right in the middle zone of the $E_{\rm p,i}$ vs. $E_{\gamma, \rm iso}$ plot validating sample. Note also that this energy corresponds to a typical GRB, being $\sim 40$ times smaller than the corresponding $E_{\gamma, \rm iso}$ inferred for the extreme case of GRB\ 080916C \citep{Abdo, Greiner}, the most energetic GRB known to date.

Since the presence of a jet break suggests a conical GRB geometry, the true gamma-ray energy release, $E_{\gamma}$\ , is smaller than $E_{\gamma, \rm iso}$\ , with $E_{\gamma} = f_{\rm b}\ E_{\gamma, \rm iso}$ \citep{Frail, Sari99}. For the above choice of parameters, computing the beaming fraction, $f_{\rm b} = 1 - \cos \theta$, with the $E_{\gamma, \rm iso}$ corresponding newly calculated angle ($\theta\sim 3.2\degr$), results in $E_{\gamma} = 1.45\times10^{50}$\ erg. Two comments can be made at this point: 

1) This rather low total energy suggests that the relatively large redshift of $ z = 5.2$ cannot be overly large.

2) This value is smaller than the average of those reported in earlier surveys, clustering around the GRBs ``standard'' energy reservoir (e.g. \citealp{Frail}). As observed later by \citet{Ghirlanda} in more recent and larger GRB samples this clustering is in fact not so narrow, covering about 2 orders of magnitude. 

Finally it would be interesting to see if GRB 050502B obeys the $E_{\rm p,i} - E_{\gamma}$ Ghirlanda relation \citep{Ghirlanda, Ghirlanda07}, but the large uncertainties in both the measured parameters and the model prevent a meaningful comparison.



\section {Conclusions}

As a main conclusion the data favor the high redshift nature of GRB 050502B. While even a very red color is not sufficient to claim a high redshift (see, e.g., GRB 060923A, GRB 070306), in this case an extra constraint is provided by $\beta_{\rm OX}$ being consistent with $\beta_{\rm X}$.
Also there is no significant excess $N_{\rm H}$ and furthermore the post-jet break slope $\alpha_{\rm X}$ is consistent with $\alpha_{\rm opt}$, again favoring a single component origin for the two bands.
The consistency of the time indices favors the other main conclusion: the existence of an achromatic jet break.

The scarcity of optical data limits a deeper discussion regarding the late engine model as an explanation for the X-ray giant flare and later activity \citep{Falcone} and likewise in considering alternative models. It is interesting to note however, that the jet break appears after the end of the last bump, which could also be originated by later engine activity. No transition is observed from the shallow phase to a pre-jet break phase, likely because the extra ``late prompt" component is added to the fireball model component. As the effects of the late engine activity cease, the afterglow shows then a standard jet break.

Given the high-redshift nature of GRB 050502B, its consequent afterglow red $R-I$ color and the fact that typical X-ray flares emit most of their flux in the $\gamma$-ray and X-ray bands (e.g. \citealp {Page,Kruehler09}), it is not very surprising that the simultaneous UVOT observation do not show evidence for a strong rebrightening in the optical wavelength regime contemporaneous with the 
prominent X-ray flare.

Finally it is interesting to note that for $z = 5.2\pm 0.3$, this would be one of the highest redshift  observed achromatic breaks, when comparing with the samples of \citet {Liang} and \citet {Ghirlanda07}.

\begin{acknowledgements}

E.P. acknowledges financial support from contracts PRIN INAF 2006 and ASI I/088/06/0. The Dark
Cosmology Centre is funded by the Danish National Research Foundation. T.K. acknowledges support by the DFG cluster of excellence ``Origin and Structure of the Universe''. This work made use of data supplied by the UK Swift Science Data Centre at the University of Leicester. We thank G. Szokoly, D. Burlon and G. Ghirlanda for useful discussions and V. Sudilovski and K. Dzurella for improvements in the draft.

\end{acknowledgements}



\bibliographystyle{aa} 
\bibliography{grb050502b} 

\end{document}